\documentclass[pdftex]{article}

\usepackage{arxiv}
\usepackage[utf8]{inputenc} 
\usepackage[T1]{fontenc}    
\usepackage[colorlinks, linkcolor=darkblue, citecolor=darkblue, urlcolor=darkblue]{hyperref}       
\usepackage{url}            
\usepackage{booktabs}       
\usepackage{amsfonts}       
\usepackage{nicefrac}       
\usepackage{microtype}      
\usepackage{graphicx}
\usepackage[natbibapa]{apacite}
\usepackage{doi}
\usepackage{booktabs}
\usepackage{color, colortbl}
\usepackage[longtable]{multirow}
\usepackage{graphicx}
\usepackage{lscape}
\usepackage[]{mdframed}
\usepackage{longtable}
\usepackage{multirow}
\usepackage{ctable} 
\usepackage{setspace}
\usepackage{array}
\newcolumntype{L}[1]{>{\raggedright\let\newline\\\arraybackslash\hspace{0pt}}m{#1}}
\usepackage{tabularray}
\usepackage{float}
\usepackage{placeins}
\definecolor{darkblue}{rgb}{0.0, 0.0, 0.55}

\title{Positive AI: Key Challenges in Designing Artificial Intelligence for Wellbeing}

\author{{Willem van der Maden} \thanks{Corresponding author: w.l.a.vandermaden@tudelft.nl}
\\
	Department of Human-centered Design\\
	Delft University of Technology\\
	Delft, The Netherlands  \\
	\texttt{w.l.a.vandermaden@tudelft.nl} \\
    \And
    {Derek Lomas} 
    \\
	Department of Human-centered Design\\
	Delft University of Technology\\
	Delft, The Netherlands  \\
	\texttt{j.d.lomas@tudelft.nl} \\
    \And
	{Malak Sadek}
 \\
	Dyson School of Design Engineering\\
	Imperial College London\\
	London, UK  \\
	\texttt{m.sadek21@imperial.ac.uk} \\
    \And
	{Paul Hekkert} 
 \\
	Department of Human-centered Design\\
	Delft University of Technology\\
	Delft, The Netherlands  \\
	\texttt{p.p.m.hekkert@tudelft.nl} \\
}

\renewcommand{\shorttitle}{Positive AI: Key Challenges}

\hypersetup{
pdftitle={Positive AI: Key Challenges for Designing Wellbeing-aligned Artificial Intelligence},
pdfsubject={cs.AI},
pdfauthor={Willem van der Maden},
pdfkeywords={First keyword, Second keyword, More},
}

\begin{document}
\maketitle


\begin{abstract}
Artificial Intelligence (AI) is a double-edged sword: on one hand, AI promises to provide great advances that could benefit humanity, but on the other hand, AI poses substantial (even existential) risks. With advancements happening daily, many people are increasingly worried about AI's impact on their lives. To ensure AI progresses beneficially, some researchers have proposed “wellbeing” as a key objective to govern AI. This article addresses key challenges in designing AI for wellbeing. We group these challenges into issues of modeling wellbeing in context, assessing wellbeing in context, designing interventions to improve wellbeing, and maintaining AI alignment with wellbeing over time. The identification of these challenges provides a scope for efforts to help ensure that AI developments are aligned with human wellbeing.
\end{abstract} 
\section{Introduction}

The rapid advancement and adoption of generative AI (GenAI) technologies like \textit{ChatGPT }signify the dawn of “The Age of AI.” \citep{gates2023age,kissinger2021age} These developments mark a significant leap in the capabilities and adoption of AI systems. However, for many people, the swift and disorienting integration of AI into daily life raises many issues \citep{cugurullo_fear_2023,fietta_dissociation_2022,qasem_chatgpt_2023}. Concerns include the potential impacts on employment, privacy, and inequality, along with broader societal implications like human rights, mental health, and the preservation of democratic norms \citep{future_of_life_institute_pause_2023, prabhakaran_human_2022, stray_aligning_2020, shahriari_ieee_2017}. This article argues for the importance of wellbeing as a key objective in AI and for human-centered design (HCD) as a key methodology. Based on this framing, it shares a set of key challenges that will face designers of AI for wellbeing, or \textit{Positive AI}. 

The idea that AI should support wellbeing is not uncommon. In 2018, \citet{zuckerberg_facebook_2018} (CEO of Meta, previously Facebook) publicly stated that wellbeing should be the goal of AI. Further, in an interview Jan Leike \citep{wiblin_jan_nodate} (head of the `Superalignment’ research lab at OpenAI) said AI optimization should align to “flourishing.” Wellbeing, however, is complicated. It is not a naturally observable quantity, but rather a multifaceted construct that is based, at least in part, on conscious human experiences \citep{ruggeri2020well}. Therefore, designing \textit{Positive AI }requires understanding and shaping human experiences. This situates the challenge squarely in the domain of human-centered design \citep{auernhammer_human-centered_2022}. Before reviewing the possibilities for HCD in designing AI for wellbeing, we will briefly address other fields associated with the creation of positive human outcomes in AI. The current article is not the venue for reviewing them in-depth. Yet, we find it important that Positive AI designers are aware broadly aware of their contributions. 

\subsection{Ethical AI}

AI ethicists have been formulating ethical principles and frameworks to responsibly guide the development and implementation of AI systems. A key contribution came from \citet{floridi_ai4peopleethical_2018} who synthesized a set of core ethical principles like beneficence, non-maleficence, autonomy, justice, and explicability for cultivating a “Good AI Society.” However, as principles alone they are insufficient \citep{mittelstadt2019principles}: they need to be translated into concrete practices. Much work remains to develop these into practical tools and methodologies \citep{morley_what_2020}. Recent work has begun exploring approaches for embedding ethical values directly into AI system design \citep{klenk_ethics_2021, van_de_poel_embedding_2020}, from which Value-sensitive Design (VSD) has emerged as a candidate to bridge the principle-practice gap \citep{umbrello_mapping_2021}. A recent critical review \citep{sadek2023designing} indicates that VSD may be effective but limited. Some limitations include the inadequate elicitation of values, a tendency to depend on pre-established values over context-specific ones, and a lack of precise instructions for embedding values.

\subsection{AI Alignment}
Considering the potential for harm done by AI, some refer to such systems as misaligned with human values. For example, referencing social media, ethicist Tristan Harris says that by optimizing for attention, these platforms are misaligned with human wellbeing and dignity \citep{harris_2017}.  “AI alignment” is a field of research that aims to develop systems that are aligned with human values and intent \citep{christian_alignment_2020}. Alignment has been earlier studied as the principal-agent problem in economics and law, where an agent must achieve the objectives and interests of the principal \citep{hadfield-menell_incomplete_2019}. For example, in a car repair scenario, the car owner (principal) expects the mechanic (agent) to fix the car efficiently and affordably. However, the mechanic might suggest unnecessary repairs to increase the bill (misalignment), contrary to the owner's desire for cost-effective service. Taking this framework to AI systems, alignment means ensuring that AI agents effectively and reliably pursue the goals and preferences set by their designers and users. One successful example of technical alignment work is the use of Reinforcement Learning with Human Feedback (RLHF) \citep{christiano2017deep}, which uses human preference data to align the behavior of Large Language Models (LLMs). Related techniques include Constitutional AI \citep{bai_constitutional_2022} and inverse reinforcement learning (IRL) \citep{ng2000algorithms}. There are many new techniques in the expanding field of technical AI alignment.\footnote{For a comprehensive overview, readers are referred to the live agenda summarizing ongoing alignment efforts posted to the AI Alignment Forum \citep{technicalities_shallow_2023}, which has been founded by prominent alignment researcher Eliezer Yudkowsky.} However, while these technical efforts show tremendous progress, their technology-centered perspective risks missing broader sociotechnical considerations, such as the design of human systems to effectively respond to AI \citep{dung2023current}.

\subsection{Why human-centered design?}

Given the intrinsic relationship between wellbeing and conscious experience, some scholars have argued for the importance of human-centered design (HCD) in AI \citep{calvo_positive_2014, desmet_positive_2013}. One reason is that, as a field, HCD focuses on understanding and shaping human experiences. However, there are a variety of ways in which Human-centered design (HCD) might complement ethical perspectives and address gaps in the AI alignment field. For instance, HCD might help bring concrete implementation methods and a broader systemic perspective. A core tenet of HCD is to prioritize the needs, values, and capabilities of users, ensuring that the design process is centered around human beings and their interactions with technology. 

 Designers are trained to attend to---and empathize with—human experiences \citep{norman2013design}. This means considering the full context surrounding users and technologies, rather than just narrow functionality, as well as prioritizing the understanding of diverse users' needs and experiences from their point of view \citep{sanders2008co}. They are equipped with the ability to engage in stakeholder participation and reveal the ethical priorities and deeply-held beliefs relevant to design projects \citep{zhang2023stakeholder}. This encompasses a blend of competencies from engineering design, including problem definition, scoping, and rapid prototyping, combined with methodologies from social sciences like conducting ethnographic research, interviews, deriving understanding from qualitative data, and engaging in empathetic practices \citep{kramer_characterizing_2016}.

These skills are particularly important for AI because the integration of diverse perspectives ensures that both technical efficiency and societal impacts are considered in AI development \citep{auernhammer2020humanb}. For instance, experimentation and prototyping in AI benefit from this blend, allowing for iterative refinement and alignment with human needs and values. Prototyping in AI can be difficult because of the inherent unpredictability and complexity in AI's capabilities and outputs \citep{yang2020re}. HCD may help by applying user-focused approaches to manage these uncertainties. Moreover, involving end users directly in the design process ensures that AI solutions are tailored to real-world requirements, making the technology more accessible, usable, and effective \citep{li2021engaging, zhu2018explainable}. In summary, the HCD perspective can complement existing ethical and technical viewpoints in AI development, as it offers methodologies to create systems that balance technical robustness with socially responsible outcomes that benefit people and society at large.

The field of Positive Design focuses on promoting human flourishing. The Positive Design Framework provides a scaffold for solutions that can enhance subjective wellbeing through components like pleasure, meaning, and virtue \citep{desmet_positive_2013}. Grounding positive design in theory and evaluating its effect through controlled studies helps ensure that designed solutions truly contribute to people's happiness. Similarly, the Positive Computing \citep{calvo_positive_2014, calvo_design_2019, gaggioli_positive_2017} movement aims to leverage technology to measurably improve wellbeing and human potential. The emphasis on collaborations between fields like psychology, computer science, and design in positive computing underscores the importance of an interdisciplinary, human-centric approach for developing AI focused on wellbeing objectives \citep{calvo_editorial_2016}. In many ways, the tenets of positive design and positive computing have helped lay the foundation for what we now call “Positive AI.” 

\subsection{Why wellbeing?}
A growing movement of scholars advocates for the incorporation of wellbeing metrics into AI systems so that optimization efforts can measurably contribute to social benefit \citep{schiff_ieee_2020, shahriari_ieee_2017}. Specifically, they argue that measures of wellbeing can help manage AI's effects on society \citep{musikanski_artificial_2020}. Indeed, wellbeing has a strong methodological foundation \citep{stray_aligning_2020}, and there is extensive research on defining and measuring wellbeing; this suggests that algorithmic systems may be able to systematically optimize wellbeing \citep{havrda_2020_enhanced}.

Wellbeing's complexity captures many relevant societal concerns AI systems should address \citep{stray_aligning_2020}. This combination of rich meaning and inherent measurability supports the operationalizing wellbeing as an optimization objective for AI systems. This sentiment is also expressed by a recent IEEE standards review that argues for the adoption of holistic wellbeing frameworks (like IEEE 7010) to guide the design, deployment, and evaluation of AI systems \citep{schiff_ieee_2020}. However, significant questions remain regarding whether available wellbeing frameworks are fully sufficient, whether existing metrics are sufficient, what the impacts of wellbeing optimization may be \citep{musikanski_artificial_2020, schiff_ieee_2020, stray_aligning_2020}.

Some argue that wellbeing is a sort of ultimate objective: in \textit{The Moral Landscape}, \citet{harris_moral_2010} argues that other values like fairness, transparency, or accountability should be seen as components that contribute to wellbeing, rather than ends in themselves. From this perspective, optimizing for wellbeing involves optimizing for all values that matter, but only insofar as they contribute empirically to wellbeing. In so far as AI systems are able to assess their own impact on human wellbeing, they may be able to potentially maximize all benefits and minimize all harms experienced by users and society \citep{havrda_well-being_2023}. Wellbeing optimization might then allow for the management of complex issues like misinformation and inequality associated with AI systems \citep{stray_aligning_2020}.

\subsection[Framing the challenges]{Framing the challenges: human-centered design of AI systems}
As a term, ‘Artificial intelligence’ is used to describe both a characteristic of computer systems and the methods employed to develop this feature, such as machine learning (ML) \citep{gabriel_artificial_2020}. Intelligence in both humans and machines has been defined as “an agent’s general ability to achieve goals in a wide range of environments.” \citep[p. 9][]{legg_collection_2007} Following this definition, AI researchers Stuart Russel and Peter Norvig define \textit{artificial }intelligence as a \textit{designed }agent that perceives its environment through sensors and acts upon that environment using actuators \citep{russell_artificial_2022}. The result of these sensors and actuators is a feedback loop that incorporates system output (e.g., action in its environment) as input for its future actions (e.g., the action had the desired effect). A cybernetic perspective examines these broader feedback loops between AI systems, their environment, and the social context in which they operate. Thus, AI \textit{systems} (in contrast to AI/ML algorithms) can be viewed as sociotechnical systems embedded within a complex network of feedback loops \citep{van_de_poel_embedding_2020}. This broader and more systemic view of AI has been proposed as an approach to deal with some of the challenges of current and future AI systems \citep{dobbe_hard_2021, krippendorff_uncritical_2021, vanderMaden2022DesignWellbeing}.

A human-centered design perspective enables designers to look beyond “the algorithm” to consider how AI interacts within a network of social, ethical, cultural, and political factors \citep{van_de_poel_embedding_2020}. This means considering how AI influences human behavior, societal norms, and institutional structures—and how AI is, in turn, influenced. This perspective requires engaging with diverse stakeholders to understand their values and needs iterative and reflective design processes that continually assess and respond to these complex dynamics \citep{sadek2023designing}.

This perspective creates new affordances for the design of AI for wellbeing. Shaping the impact of AI can occur through multiple components of the AI system, including the technical artifacts, institutions, practices, etcetera---all in addition to the design of algorithms. For example, consider the role of AI in a video streaming platform like Netflix. A focus on the algorithm is limiting because one can only affect the likelihood of a particular recommendation. In contrast, a broader perspective opens up different kinds of interventions. For instance, in the user interface (e.g., autoplay); the organizational level (e.g., establishing boardroom content acquisition metrics beyond just engagement and growth); data science (e.g., introducing new metrics for optimization that prioritize suggesting content from more diverse voices); or the broader ecosystem (e.g., funding initiatives to broaden representation in the creative industry talent pipeline). In other words, design interventions can occur at multiple levels of the AI \textit{system}, not just in the algorithm. Designers can even consider interventions outside of the control of the AI platform, such as a ‘Netflix Watch Club’ or an alternative YouTube user interface for education \citep{lukoff2023switchtube}. 

Why is this broadened view important? Rather than aligning AI to the needs of society, there may be many cases where it may be more appropriate for social institutions to adapt to AI. For instance, while ChatGPT could potentially be aligned with the needs of K12 schools (i.e. so that students are prevented from cheating on their assignments), designers may wish to create new guidelines for positive integration of AI in their courses (e.g., promoting AI literacy in using ChatGPT\citep{mollick_meet_2023}). With this systemic perspective, there are expanded opportunities for guiding AI impacts beyond the algorithmic design itself. This shows how there are new opportunities for creating AI for Wellbeing beyond what is typically the scope of AI alignment or ethical AI research.

In this article, we conceptualize AI as a sociotechnical system involving a complex interaction of various feedback loops, each optimized for specific objectives within the system. The essence of `Positive AI' lies in harmonizing these objectives towards a singular, overarching aim: enhancing wellbeing. 

\section{Key Challenges}

As experienced design researchers in this domain, we have consistently encountered unique challenges in designing \textit{Positive AI}. The challenges we outline here are intended to inform and guide other designers embarking on similar ventures. This conceptual framework aids in structuring the challenges of designing AI for wellbeing around the following questions:

\begin{enumerate}
    \item \textbf{Modeling} the state of the system: How do we operationally define wellbeing within the context of a particular sociotechnical system?
    \begin{itemize}
        \item For example: \textit{What wellbeing dimensions are important in the context of Netflix and how do we attribute changes in wellbeing to components of the system?}
    \end{itemize}
    \item \textbf{Assessing} the state of the system: How do we translate qualitative experiences into assessment metrics?
    \begin{itemize}
        \item For example: \textit{How can we elicit how people feel about their interactions with TikTok, and how can these experiences be translated into metrics that can be used for assessing future interventions and optimization processes? }
    \end{itemize}
    \item \textbf{Designing} system actuators: How do we design interventions in AI systems that promote and enhance wellbeing?
    \begin{itemize}
        \item For example: \textit{How do we know where in the sociotechnical system of ChatGPT we should and can intervene, and how do we know whether our potential interventions will achieve the desired effect?}
    \end{itemize}
    \item \textbf{Optimizing }the system objective: How do we know whether we are getting close to our desired goal?
    \begin{itemize}
        \item For example: \textit{How do we manage tradeoffs between autonomy and social connection in designing for wellbeing on Reddit, and how do align immediate outcomes with long-term wellbeing goals? }
    \end{itemize}
\end{enumerate}

\begin{figure}[!htb]
    \centering
    \includegraphics[width=0.65\linewidth]{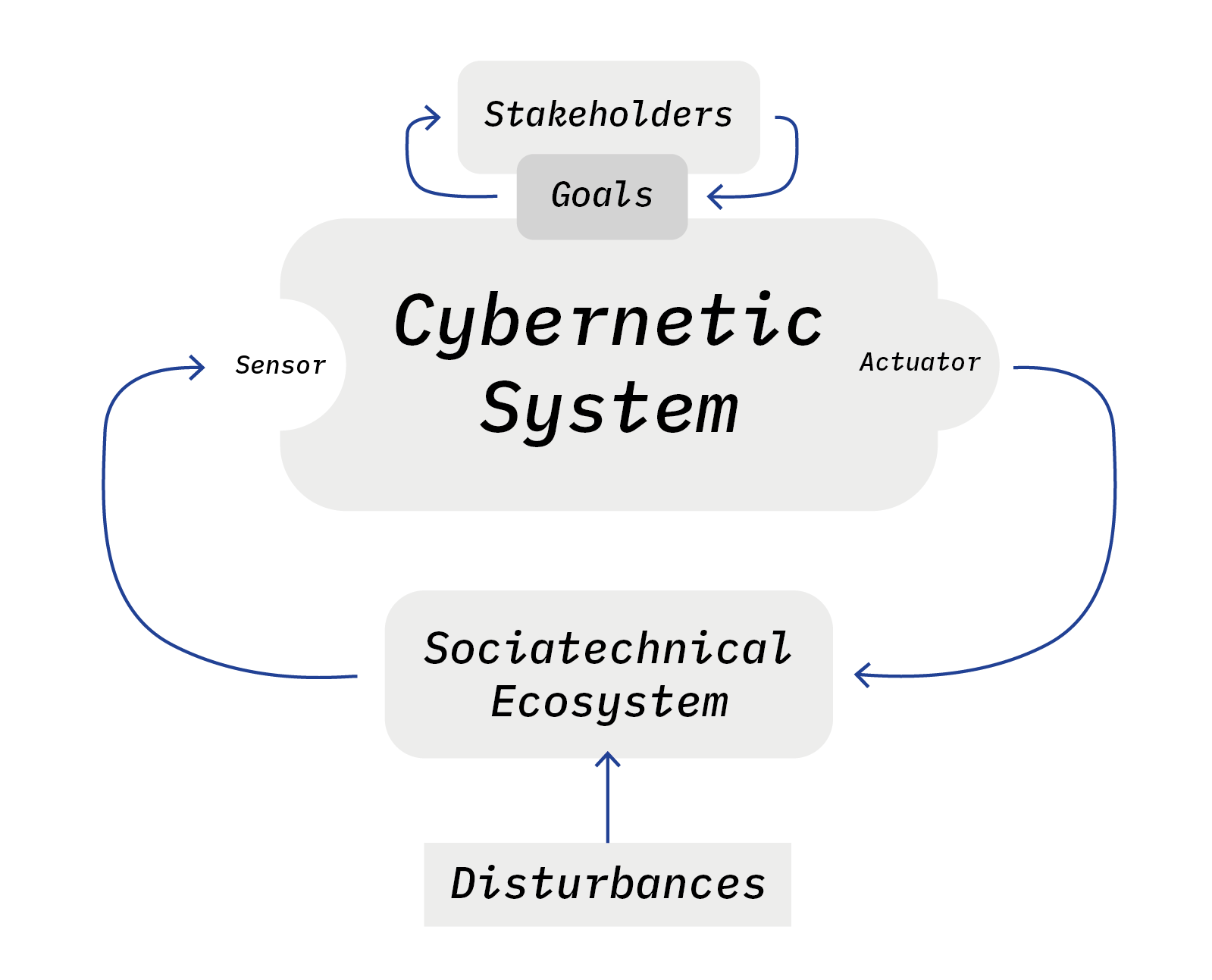}
    \caption{Shows a schematic representation of a cybernetic system. The different challenges can be mapped onto this framework: (1) understanding the system context which entails modeling the relation between wellbeing of the systems constituents and its various components; (2) operationalizing said model of wellbeing; (3) designing interventions to actively promote operationalized model of wellbeing; and (4) retaining alignment with the overall goal. The latter refers to both challenges of algorithmic optimization as well as scrutinizing the objective (e.g., is the wellbeing objective still aligned to needs and desires of all relevant stakeholders?)}
    \label{fig:cybsyschallenges}
\end{figure}

We have found it useful to use the concept of cybernetics as a lens for organizing the challenges of designing AI for wellbeing \citep{krippendorff_uncritical_2021, sato1991ai, dobbe_hard_2021}. Cybernetics allows for a systemic and holistic viewpoint that defines clear mechanisms for impact \citep{tabari_role_2022}. A cybernetic viewpoint, for us, naturally accommodates ecological and sociotechnical perspectives on the many feedback loops governing human systems and AI systems in society today. This cybernetic perspective allows us to organize the key challenges of designing AI for wellbeing into four main categories, as shown in Table \ref{tab:challenges}. This table serves as an overview and guides the structure for the rest of this chapter, with the numbers corresponding to the categories depicted in Figure \ref{fig:cybsyschallenges}.

\begin{landscape}
\begin{table}
\caption{Table 2. Overview of Key Challenges}
\label{tab:challenges}
\begin{tabular}{@{}L{2cm}L{4cm}L{10cm}@{}}
\toprule
\textbf{Category}                        & \textbf{Challenge}                                                    & \textbf{Explanation}                                                                                                                                                                                                              \\ \midrule
How to model wellbeing? & (1a) Identifying relevant wellbeing dimensions               & There is a long tradition of wellbeing research spawning many different theoretical paradigms. In any given context, the appropriate theoretical paradigm may vary.                                                      \\
                                & (1b) Modeling wellbeing and interpreting fluctuations        & The multifaceted nature of wellbeing and the richness of the world in which we live, make it difficult to model fluctuations in wellbeing to (specific features of) a system.                                            \\ \midrule
How to assess wellbeing?        & (2a) Operationalizing wellbeing                              & In order for a system to respond properly and timely, we have to measure wellbeing in a dynamic, continuous manner that is sensitive to the context.                                                                     \\
                                & (2b) Translating qualitative experiences into system metrics & We currently lack methods for connecting small-scale qualitative research on wellbeing to large-scale, quantitative methods, making it difficult to ensure that metrics of wellbeing are aligned with human experiences. \\ \midrule
How to design for wellbeing?    & (3) Designing interventions to promote wellbeing             & It is not straightforward how an algorithm, the UI of a platform, or its content may lead to wellbeing effects. There’s both a lack of examples and a lack of appropriate design methodologies for Positive AI.          \\ \midrule
How to optimize for wellbeing?  & (4a) Optimization tradeoffs and prioritization               & Optimization causes tradeoffs that are difficult to measure, making it difficult to balance different goals and predict the effects of novel metrics on the optimization process and outcomes.                           \\ 
                                & (4b) Pace layers                                             & Changes in wellbeing are slow but AI optimization cycles are fast, making it difficult to optimize for wellbeing.                                                                                                        \\ \bottomrule
\end{tabular}
\end{table}
\end{landscape}

\subsection{Challenges related to modeling wellbeing}
\subsubsection{Identifying relevant wellbeing dimensions}
Since the rise of the Positive Psychology movement that gained popularity around the turn of the millennium \citep{seligman2019positive} scientific interest in wellbeing has proliferated. Wellbeing is a multifaceted, rich, and dynamic phenomenon, and as such, there are many definitions of it in both scholarly and public contexts (e.g., policymaking). Each definition pertains9 to different aspects of wellbeing. For instance, the WHO defines health as “a state of complete physical, mental, and social wellbeing, and not merely the absence of disease or infirmity.” \citep{callahan_who_1973} There are also scholarly traditions that call wellbeing “Happiness” which can be distinguished in episodic (“I feel happy”) and attributed happiness (“I am a happy person”) \citep{veenhoven_journal_2014}. These definitions are akin to \textit{hedonic} and \textit{eudaimonic} understandings of wellbeing \citep{ryan_happiness_2001}---which are the two categories typically used to describe wellbeing research. 

Contemporary hedonic traditions tend to focus on the degree to which people experience positive and negative emotions and how satisfied they are with their life. Two reviews argue that the most prominent theoretical approach in this tradition is the tripartite model of Diener \citep{cooke_measuring_2016, linton_review_2016}. This model is typically measured in terms of “life satisfaction” \citep{diener_satisfaction_1985} and the presence of positive and absence of negative emotions \citep{watson_development_1988}.

In contrast, eudaimonic wellbeing is based on the pursuit of virtue, striving to become the best version of oneself and developing one’s personal strengths \citep{deci_hedonia_2008}. Psychologists investigating this phenomenon tend to define it in a multifaceted way, such as Ryff’s Psychological Wellbeing (PWB) scale \citep{ryff_happiness_1989, ryff_structure_1995} and Seligman’s PERMA model \citep{seligman_flourish_2010}. These theories overlap to a great extent and encompass facets such as positive relations, meaning in life, and personal growth. 

Further, there is the tradition of Quality of Life (QoL), which is often used interchangeably with wellbeing. However, the literature on the subject often pertains more to social aspects of the phenomenon and, for example, situations towards end of life, living with a disability, or life after a clinical visit \citep{bakas_systematic_2012, felce_quality_1995, moons_critique_2006}.

A final related concept is wellness, which emphasizes a holistic lifestyle (e.g., nurturing emotional and spiritual intelligence) \citep{corbin_toward_2001}. In identifying the relevant dimensions of wellbeing, a designer could also investigate non-Western traditions of defining wellbeing such as Ubuntu \citep{hailey2008ubuntu}, Ikigai \citep{sone2008sense}, and Gross National Happiness (GNH) \citep{ura_gnh_2012}.

Researchers criticizing the Western viewpoint in wellbeing science also argue that current mainstream approaches tend to be too contextualized \citep{mead_moving_2021}. Historically, psychological research on wellbeing has focused on psychometrics in order to statistically validate generalizable dimensions of wellbeing \citep{searle_introduction_2021}. However, this approach does not necessarily translate to actionable insights. Recognizing this, scholars have developed domain-specific theories of wellbeing for areas like work \citep{clifton_wellbeing_2021} and education \citep{konu_well-being_2002}. 

Contrary to global conceptions of wellbeing, domain-specific theories focus more on aspects that are prevalent in that domain which may not translate to other domains. By contextualizing wellbeing research, the findings become more directly relevant for decision-making such as in policy-making. For example, they can reveal nuances around how wellbeing manifests with a given domain such as stress experience around work relations. However, increased specificity comes at the cost of reduced situational-consistency. This shift underscores the importance of balancing context-sensitive perspectives with more generalized insights that hold over time and place. Both remain indispensable for a comprehensive understanding of wellbeing.

The concept of digital wellbeing illustrates a similar balance. Digital wellbeing being refers to wellbeing effects that typically occur in a digital context—i.e., while gaming, surfing the web, or interacting on social media. It goes beyond the mere time spent online, focusing on how digital engagement affects daily life and emotional wellbeing. It is about achieving a harmonious interaction with technology, where the benefits are maximized, and drawbacks such as loss of control are minimized. Furthermore, the IEEE 7010-7020 \citep{ieee_sa_ieee_2020} initiatives represent a pioneering effort to establish a comprehensive understanding of wellbeing within the realm of AI. Other researchers have explored the relationship between AI and community wellbeing \citep{musikanski_artificial_2020} investigating the development of wellbeing metrics, community-centric AI, and applying AI to enhance community wellbeing. 

Wellbeing, as highlighted in various perspectives, presents designers of AI systems with a range of dimensions and paradigms to consider. The relevance of wellbeing aspects varies by context; for example, physical health may be central in diet apps, while social media platforms might emphasize social connection. This complexity in people's interactions with technology is highly important for designers to consider. It requires careful, context-specific selection of the most relevant wellbeing paradigms and domains. Currently, there is no agreed-upon process for determining the which wellbeing dimensions to prioritize for a given system. This complexity in people's interactions with technology, being situational and dynamic, is reflective of findings in the literature \citep{vanden_abeele_digital_2021}. As AI systems become increasingly entangled with daily routines, modeling their impacts on wellbeing becomes difficult. Thus, selecting appropriate theoretical dimensions is crucial yet challenging. Designers, therefore, face the complex task of modeling wellbeing in AI systems within evolving sociotechnical landscapes, a challenge we will explore further.

\subsubsection{Modeling and attributing wellbeing changes}

For AI to effectively promote wellbeing, it requires a deep understanding of what actions and strategies contribute to this goal. This, in turn, requires a contextual model of wellbeing that allows the designer to attribute fluctuations to specific features of the system. Contextual models of wellbeing explain how various aspects of wellbeing manifest situationally and connect to components of a given system. The discussion below focuses on the challenges in developing contextual wellbeing models and attributing wellbeing fluctuations to components of that model, starting with the former.

Considering the complexity of AI systems and contexts, it is crucial to develop an understanding of how different aspects of wellbeing relate to specific features or components of a particular context. Quantitative measures enables researchers to objectively examine explanations and predictions from conceptual models addressing the determinants and outcomes of wellbeing \citep{diener_assessing_2019}. However, developing such models poses significant challenges. The relationships between various causal factors and wellbeing remain unclear, complicated by bidirectional and nonlinear effects. For example, while good sleep benefits mental health, improved mental health also leads to better sleep \citep{scott_improving_2021}. Similarly, the connection between mental health and social media use is complicated: it is unclear whether people turn to social media when they are already struggling with mental health issues, or whether social media itself can contribute to these issues \citep{coyne_does_2020, hjetland_pupils_2021}.

The open questions around the relation between specific media and wellbeing have implications for the deployment of AI technologies aimed at promoting wellbeing \citep{johannes_no_2022}. Specifically, they highlight the need to consider the long-term impacts of these technologies on individuals’ psychological health and wellbeing—e.g., watching Netflix may be conducive to wellbeing in the current moment but how may it shape effects over time? Thus, to design Positive AI, it is necessary to have a thorough understanding of the relationships between wellbeing and its various antecedents. Without this understanding, it is challenging to measure the impact of interventions and determine their effectiveness. The real world is characterized by complex, interconnected patterns that can make it difficult to attribute changes in wellbeing to a single event or intervention \citep{fokkinga2020impact}. This is particularly true when dealing with “narrow” AI systems, such as recommendation algorithms that suggest products a user might like based on their past preferences. These systems may excel at completing specific, limited tasks but struggle to account for the full range of factors that can impact wellbeing in the real world. For example, a product recommendation system would not consider how using that product might affect a user’s sleep, relationships, or long-term wellbeing. 

This modeling poses difficulties not just due to AI system opacity \citep{gabriel_artificial_2020}, but because any platform comprises only a fraction of a person's broader life experience. For example, changes in the dietary practices of a teenager may be a result of the content of their Instagram feed (e.g., only images of people with a specific body type) or because of some other event that occurred in their life (e.g., a breakup or the start of a new fitness program). In order to establish causality, wellbeing has to be measured in a manner that considers the complexity of the context in which the system is deployed. This poses a challenge because existing wellbeing assessments and platforms are not designed to consider the complexity of people’s lives and experiences that extend beyond the platform. For instance, how might Instagram account for life events that occur outside of its platform and influence a user’s wellbeing? And how would it determine which of those external factors are most relevant and impactful? The question of whether Instagram should be held accountable for the wider impact of its user interactions on wellbeing is a matter of ongoing debate. However, for the platform to effectively influence wellbeing, it is imperative that it addresses these interaction effects in some capacity.

In summary, developing contextual models that effectively attribute fluctuations in wellbeing to AI systems poses profound challenges. The interconnected relationships between wellbeing and other life factors resist straightforward causal analysis. This difficulty intensifies for narrow AI platforms, as they comprise limited slices of broader human experience. Although difficult, advancing more contextualized and systemic modeling methodologies promises significant progress toward the goals of Positive AI.

\subsection{Challenges related to assessing wellbeing}

Wellbeing is suited for AI optimization given its history of measurement \citep{stray_aligning_2020}. However, to measure wellbeing in context, we need to model wellbeing in context. For this, we need qualitative inputs to understand subjective personal and community experiences. Translating qualitative insights into quantitative metrics usable for optimization is non-trivial. Here, we discuss challenges related to effective assessment, which requires contextualization and bridging gaps between individual perspectives and system-level scales.

\subsubsection{Contextually operationalizing wellbeing}

As previously discussed, in order to attribute fluctuations in wellbeing to components of a given system, wellbeing must be effectively modeled and measured. While we have explored the theoretical challenges associated with this task, there are also methodological aspects that need to be addressed. This includes the operationalization of chosen wellbeing dimensions.

Wellbeing has traditionally been measured in field of psychology, but since the last decade, there has been an increased interest in also measuring wellbeing for other purposes, such as policymaking \citep{frijters_handbook_2021}. The assessment of wellbeing is often done using qualitative surveys and interviews \citep{alexandrova_well-being_2012}. It is important that the assessment instruments employed in these studies are validated, possess temporal stability, and demonstrate cross-situational consistency \citep{diener_assessing_2009}. However, literature suggests that these “off-the shelf” wellbeing assessment instruments may not be readily applicable in the context of novel technologies that rapidly change factors influencing wellbeing, such as social media \citep{kross_social_2021} and AI \citep{stray_building_2023}. Traditional instruments aiming for consistency over time and place may be incompatible with contexts that transform quickly and substantially. This also means that present modalities of assessment (e.g., surveys and interviews) may not be sufficient for these emerging contexts \citep{stray_building_2023}.

Given these limitations, there is a need for measures that are better suited to assess wellbeing in the context of rapidly changing technologies and situations \citep{vanden_abeele_digital_2021}. Context-sensitive measures, designed to adapt to the specific needs and context of the individuals being assessed \citep{loveridge_measuring_2020, van_der_maden_framework_2023}, provide a promising alternative and have been suggested to be more suitable for the operationalization of values (including wellbeing) in concrete applications \citep{liscio_axies_2021}. The value inference process outlined by \citet{liscio2023value} highlights the importance of identifying context-specific values as a crucial step in aligning AI agents with human values. Their methodology demonstrates an approach to elicit relevant, contextualized values that could inform the development of context-sensitive wellbeing measures. These measures allow for a more tailored assessment of wellbeing, as they can be regularly updated to reflect the changing needs of the community. It should be noted that this adaptability poses tradeoffs regarding validation and stability over time and contexts. How to reconcile the discrepancy between the need for sensitive, customized measures and generalizable instruments remains an open question warranting further investigation. 

Finally, to develop context-sensitive measures of wellbeing, it is crucial to engage regularly with the community to scrutinize and update the instruments used to assess contextual wellbeing. The literature has identified the importance of community engagement to implement ethical frameworks \citep{morley_ethics_2021} and articulate shared values in AI systems \citep{sanderson2023ai}, though sources acknowledge there are gaps in best practices for stakeholder participation \citep{sadek2023co}. While human-centered design methods may provide useful directions for establishing community engagement, they need to be adapted to the particularities of AI systems. Human-centered design excels at developing a deep understanding of contexts, maintaining community participation, and understanding individual and community needs. In conclusion, contextualizing wellbeing measures involves navigating consistency-sensitive tensions and engagement complexities. However, even well-constructed contextual measures struggle to inform AI optimization unless translated to system-level data. This underscores the pivotal challenge of connecting qualitative insights with large-scale, quantative metrics for AI alignment.

\subsubsection{Translating qualitative experiences into system metrics}

Wellbeing is a highly personal experience that typically requires individuals to report on their own experiences in order to be measured \citep{diener_assessing_2009, linton_review_2016}. As such, qualitative methods such as 1-to-1 interviews, focus groups, and ecological momentary assessments, are often expected to provide more contextually actionable data. However, currently, there is no agreed-upon process for translating small-scale activities of this nature into large-scale optimization metrics \citep{mcgregor2015competing}. Instead, easy-to-collect but incomplete or inaccurate metrics are used—such as hours spent on a social media platform to measure satisfaction with content—which do not fully capture users’ experiences and overlook harmful consequences to their wellbeing \citep{thomas_reliance_2020}. Conversely, connecting qualitative research on wellbeing with large-scale optimization methods could not only help ensure that measures of wellbeing are well-aligned with human experiences, but also help identify areas in which measures of wellbeing can be improved \citep{camfield_enquiries_2016}. Aside from issues of scalability, there exists a discrepancy between the types of metrics suitable for on-platform measurement versus those typical for wellbeing assessment.

Whereas self-reporting suggests to be the best way of measuring wellbeing, behavioral data collection is the default method for on-platform optimization. However, behavioral metrics cannot reliably measure wellbeing since research on the relationship between behavioral and self-reported wellbeing measures is limited and inconclusive \citep{dang2020self}.  Currently, it is unclear whether behavioral metrics can replace self-report measurements. Relying solely on self-reporting may, however, negatively impact the user experience, as users should not be bombarded with wellbeing questions upon engaging with a platform. This presents a challenge for designers and design researchers: how can platforms facilitate user feedback to collect accurate and scalable wellbeing data? This non-trivial challenge has also been acknowledged by other researchers \citep[e.g.,][]{steur_properties_2021}.

In conclusion, effectively integrating wellbeing requires translating between qualitative experiences and system metrics. While qualitative insights like self-reports capture personal experiences vital for alignment, convenient behavioral metrics dominate on-platform data collection. Absent mechanisms to translate small-scale activities into optimization inputs, AI risks misrepresenting user needs. Progress necessitates the development of methods that bridge this divide —continuously engaging individuals and communities while surfacing priorities at a systemic level. However, creating participatory channels poses immense practical difficulties around incentivization, standardization, and scalable synthesis. Though an open challenge, pioneering such participatory architectures in ways that meaningfully empower stakeholders promises to actualize AI’s potential for responsibly nurturing human flourishing.

\subsection{Challenge related to designing for wellbeing}
We have discussed the difficulties in conceptualizing and operationalizing wellbeing for AI systems. We have also touched on the need to optimize across individual, community and societal levels of wellbeing. A further question is: given a concept and operationalization of wellbeing, what actions can an AI system take to positively impact wellbeing? And how may we design such actions?

\subsubsection{Methodological challenge of designing (AI) actions to promote wellbeing}
Designing AI is an incredibly challenging task for designers \citep{sadek2023designing, yang2020re}. Before tackling the design of AI that promotes wellbeing, it is important to understand the difficulties that designers face when working with any AI-based system in general. Currently, the communication gap between designers, developers, and end-users \citep{yang2020re, yu2020keeping} causes an ``AI support vacuum” \citep{abaza2021} where AI neither supports stakeholders nor is supported by them. Despite calls for the more interdisciplinary design of AI-based systems \citep{harbers_towards_2022, west_discriminating_2019}, there is a lack of ‘translational work’ in current interventions that aim to support this collaborative design \citep{wong2022}.  Aside from communication-based challenges, without proper training, it is difficult for designers to understand how to ideate, design and prototype for AI-based systems \citep{allen2018}. It is important to mention and consider these challenges before examining the extra difficulties that a focus on wellbeing might present.

Following the earlier discussion of adopting a systemic perspective, it is currently unclear whether designing Positive AI requires changes to the interface, algorithms, content moderation policies, business models, or otherwise defined components. Interventions could even extend beyond the system itself. This uncertainty stems from a lack of examples and established methods for putting wellbeing at the core of AI design. For example, current strategies for promoting wellbeing through ChatGPT remain largely undefined and untested. Designing Positive AI requires fundamentally rethinking how we approach design to focus on the continuous measurement of wellbeing and alignment with human values. There are well-established frameworks for designing with wellbeing in mind, such as Positive Design \citep{desmet_positive_2013} and Positive Computing \citep{calvo_positive_2014}. These approaches acknowledge the need to consider the impacts on wellbeing, but they were not developed with the unique complexities of AI systems (such as system opacity, unpredictability, and scalability) in mind.

In addition to design methods, alignment approaches like Constitutional AI \citep{bai_constitutional_2022}, Inverse Reinforcement Learning (IRL) \citep{arora2021survey, ng2000algorithms} and Contestable AI \citep{alfrink_contestable_2022} also aim to ensure AI systems remain human-centered and aligned with human values. While these methods primarily focus on modifying the AI system’s internal processes and decision-making mechanisms, they highlight the importance of incorporating human values and wellbeing considerations directly into AI systems. By combining the strengths of these alignment approaches with design methods, we can develop a more holistic and effective AI alignment strategy that addresses user experience, interaction, and overall design, which are crucial for affecting wellbeing. For instance, some platforms, like YouTube and Twitter, have made small changes intended to benefit wellbeing, such as removing dislike counts and allowing users to `unmention’ themselves from conversations. Whether these are cases of `ethics washing’ \citep{floridi_unified_2019} or have actual benefit to wellbeing is currently unknown to the public and academia, as research on their effects is not shared publicly. 

In this vein, \citet{stray_aligning_2020} discuss Facebook’s MSI metrics and YouTube’s satisfaction metrics as cases of AI optimization aligned to wellbeing. However, they criticize that both companies did not involve or get feedback from the people affected by their AI changes. The absence of public assessment and unclear impact on broader aspects of wellbeing, such as social connectedness or life satisfaction, particularly in diverse communities, underscores a crucial shortcoming. This lack of detailed information and engagement hinders a comprehensive understanding of the effectiveness of these interventions in truly aligning AI with community wellbeing. This is a broader concern in the value-sensitive design (VSD) of AI. That is, as discussed in the background section, while VSD methods support the identification and embodiment of values for AI well, they lack support in assessing ‘realized’ values—i.e., whether the designed outcomes in fact achieve the intended effect on said value \citep{sadek2023designing}. This is particularly important for Positive AI, because the core premise lies in empirically confirming that systems positively impact human wellbeing. Without closing the loop between intended and actualized outcomes, the benefits of proposed interventions remain theoretical.

To recapitulate, designing AI actions that support wellbeing is a complex challenge that requires a combination of design methods and alignment approaches. Successful regulation and alignment requires internal diversity matching external complexity. Without design methods and alignment approaches considering the full range of variables shaping wellbeing, AI systems will struggle to effectively promote flourishing While existing design methods provide guidance, further research and development is needed. By focusing on continuous measurement of wellbeing, alignment with human values, and incorporating aspects of both design methods and alignment approaches, we can work towards creating Positive AI systems that not only avoid harm but actively promote human flourishing.

\subsection{Challenges related to optimizing wellbeing}

\subsubsection{Optimization tradeoffs}
Maximizing user engagement to drive revenue is a central optimization challenge for most platforms, where varied metrics like views, likes, and shares track user interaction with content. Using multiple metrics allows optimization algorithms to take a multi-objective approach to personalize content for optimal user engagement \citep{trunfio_conceptualising_2021}. However, balancing these objectives already involves tricky tradeoffs \citep{thorburn_how_2022}. For example, there is tension between showing novel content to pique interest and only showing content matched to the user's interests to avoid disengagement \citep{lu2020beyond}. While some novelty draws users in, too much risks boring them with irrelevant content. Now, considering wellbeing optimization makes this process even more complex.

Firstly, it is difficult to determine which facet of wellbeing should be prioritized—both in the moment and over time. The various facets of wellbeing often compete with one another. For example, social media platforms must balance users’ needs for social connection and personal autonomy. Promoting social interaction may support wellbeing by facilitating relationships, but it could also infringe on users' freedom to choose their own activities. Additionally, an individual's priorities may shift over time. What enhances wellbeing in the short-term may differ from long-term needs (e.g. enjoying frequent social activities when moving to a new city versus after settling in). This complexity requires nuanced techniques that can account for tradeoffs between competing needs and evolving individual priorities. 

Secondly, because of these optimization tradeoffs there is no optimal solution. This is a common issue in environmental sciences who face complex tradeoffs between, for example, biodiversity and human wellbeing \citep{daw_evaluating_2015, mcshane_hard_2011}. This also goes for optimizing for just human wellbeing in the sense that there is no “right” outcome in balancing, for example, the wellbeing of an individual over that of a family (who are likely to value different things). Here the notions of requisite variety and satisficing become important again. That is, as platforms are faced with these complex tradeoffs they should develop many different strategies to deal with problems as they come up and take small incremental steps to deal with tradeoffs over time. 

Thirdly, prioritizing one wellbeing facet over others may yield unintended consequences—beyond just direct tradeoffs. Focusing narrowly on a single aspect of wellbeing can backfire and undermine that very facet over time, given the complexity and interconnectivity of wellbeing. For instance, features meant to enhance social connectedness could hamper autonomy or other unforeseen needs. Wellbeing's multidimensional nature means myopic solutions risk negative ripple effects from complex causal interrelations that remain poorly understood. Therefore, maintaining broad sensitivity to these complex interactions is critical. By regularly reassessing for subtle harms and tightening feedback loops, platforms can progressively identify and resolve unintentional side effects. An open, responsive systems perspective allows more complete understanding to emerge gradually from ongoing learning. Overall vigilance to complexity and interconnectedness may better serve wellbeing than rigid prioritization of singular facets.

Therefore, when designing AI systems for wellbeing, the models guiding decisions should be continually reassessed through collaboration with relevant stakeholders. This allows for adaptive alignment as understandings of wellbeing, design contexts, and community needs evolve over time. Continual engagement enables updating system priorities to restore balance across wellbeing dimensions. However, even with continual reassessment and stakeholder collaboration, another fundamental challenge persists---the differential pace of change between wellbeing and AI optimization.

\subsubsection{Fundamental challenge of pace}
Changes in wellbeing typically manifest over longer periods of time. For instance, life satisfaction is not expected to fluctuate dramatically from week to week \citep{pavot_temporal_2014}. While it can be argued that this is due to our assessment instruments \citep{boschman2018within}, the argument remains that this is in stark contrast to the pace of AI optimization and, for instance, media consumption, which are much faster. This can make it difficult to reconcile the two, as the wellbeing effects of AI actions should inform the system’s optimization cycles. Note here that system optimization cycles here refer to both analog (e.g., managerial, and designerly) and digital (e.g., algorithmic news feed recommendations) optimization cycles. 

In other words, from the perspective of the optimization algorithm, the best it can do is optimize for hedonic wellbeing (momentary pleasure), while the goal is to design AI that supports eudaimonic wellbeing too (long-term wellbeing). An AI system can optimize for what feels good in the moment but not for what is good for you in the future. For example, watching “just one more episode” may be desirable in the moment (hedonic) but regretted the next morning when you feel tired during an important meeting (eudaimonic). Alternatively, from a managerial optimization perspective, linking executive goals to user wellbeing metrics may show progress only after longer periods than quarterly reporting cycles. Yet leaders frequently anticipate prompt, measurable outcomes that can be effectively communicated during shareholder meetings, reflecting a preference for quick, tangible results. This pace mismatch risks reactive changes before initiatives fully play out—such as prematurely disbanding a wellbeing taskforce. Evidently, there is a need to reconcile these pace layers somehow.

\begin{figure}[!htb]
     \centering
     \includegraphics[width=0.5\linewidth]{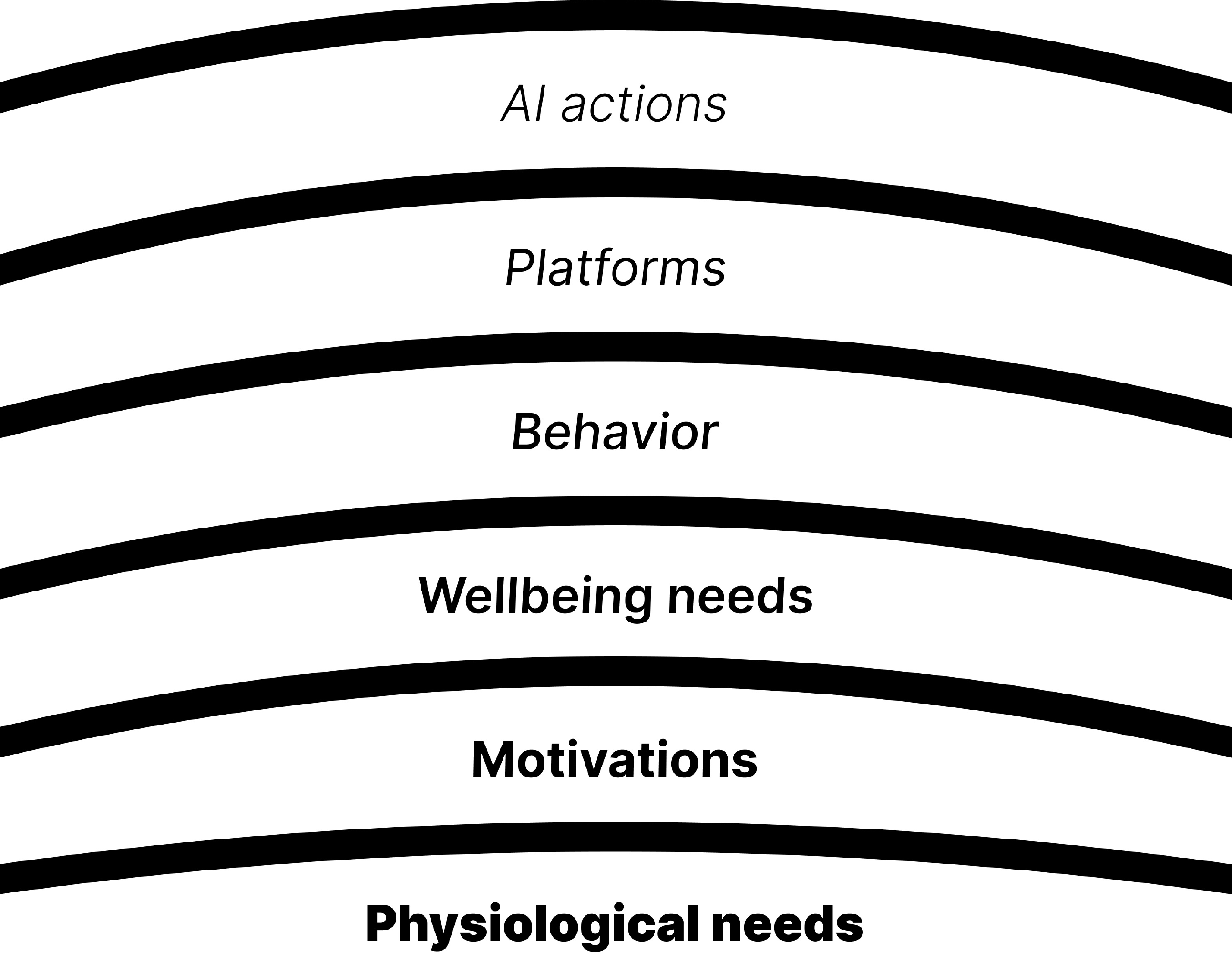}
     \caption{Different systems operate at different paces, adapted from \citet{stephen_pace_2021}}
     \label{fig:pacelayers}
 \end{figure}

Here, we can look to Stewart Brand’s theory of pace layering \citep{brand_pace_2018}, which suggests that different systems evolve at different speeds (see Figure \ref{fig:pacelayers}). The pace of platform change far outpaces that of wellbeing. This is a fundamental constraint: the pace of AI optimization processes and the timeline for observable changes in human wellbeing are mismatched, and the rate of these dynamics in either domain is unlikely to significantly change in the future. Acknowledging and making explicit that there are in fact different components in a system that evolve at different paces is essential as we should look for ways to bridge or translate across layers. That is, we need an intermediary layer for measurement—for which qualitative methods may be well-suited. In the example given before, Netflix may choose to incorporate qualitative feedback from users regarding their overall satisfaction with their viewing habits, including how it affects their daily lives, sleep patterns, and long-term goals—bridging the gap between behavior and AI actions.

\section{Discussion}
This paper aimed to outline key challenges designers face when developing AI systems to actively promote human wellbeing, termed Positive AI. It makes two main contributions. First, it proposed adopting a cybernetic, systemic perspective for conceptualizing and addressing these alignment challenges. This viewpoint emphasizes the sociotechnical nature of AI systems, considering potential interventions across multiple system levels. Second, it organized the complex issue of designing Positive AI into four main categories of challenges: 1) modeling wellbeing, 2) assessing wellbeing, 3) designing for wellbeing, and 4) optimizing for wellbeing. These categories provide structure for mapping relevant problems within a systemic framework geared towards continuous improvement through stakeholder participation.

To recapitulate the challenges, designing Positive AI is facing substantial gaps regarding our knowledge on how to do it. The abundance of theoretical wellbeing paradigms makes it difficult for designers to get started on modelling wellbeing within a given context. This leads to issues in measuring wellbeing and attributing changes across different scales, from individuals to communities to sociotechnical systems. Further, the complexity of wellbeing introduces optimization tradeoffs, raising questions around balancing competing wellbeing needs between individuals and groups over time. While these present methodological and technical difficulties, they are fundamentally design challenges requiring creative solutions.

Because of these challenges, in some cases, platforms may opt for simpler derivatives of wellbeing \citep{pan2022the}, or designs that merely mitigate illbeing or improve aesthetics, rather than investing resources in optimizing for wellbeing itself. However, the impending mental health crisis \citep{world_health_organization_world_nodate} and the transformative impact GenAI \citep{chui2023economic} make it all the more urgent to move towards Positive AI that systematically prioritizes wellbeing. This requires more than just a focus on mitigating harm; it is about capitalizing on each platform’s potential to foster wellbeing. 

The key takeaway is that adopting a cybernetic perspective that places wellbeing assessment at the core can guide this process. It emphasizes the need for design researchers to continuously re-examine contextual models of wellbeing. This requires engaging relevant stakeholders, such as end-users and developers. By framing wellbeing as an ongoing conversation, we can iteratively refine models and measurements as contexts evolve through stakeholder participation. This reflexive, adaptive approach allows designers to navigate complexity and uncertainty when developing AI systems aimed at fostering human flourishing.

\subsection{Wellbeing as an objective for AI optimization }

The concept of optimizing metrics wellbeing in AI systems emerges as a response to the limitations of traditional metrics like profit or efficiency \citep{schiff_ieee_2020, stray_aligning_2020}. While financial returns have been the primary focus for many companies, this singular pursuit may lead to broader societal harms \citep{virokannas_contested_2020}. Might wellbeing as a guiding principle for AI design serve as a compelling alternative? Wellbeing, with its quantifiable and multidimensional nature, encompasses various aspects of human life which may make it a more suitable optimization goal. It allows for concrete optimization metrics that are more aligned with human-centered values, potentially reversing the trends of societal harm caused by narrowly focused objectives.

Yet, it seems impossible (and ill-advised) to reduce wellbeing to a single metric. Instead, wellbeing should serve as a multidimensional guide for value-based decisions and a comprehensive principle for moral choices. 

Despite arguments for optimizing AI systems for wellbeing, companies may still opt for simpler derivatives due to factors like potential public relations issues, unclear financial rewards, and risks of losing competitive advantages. For example, companies may hesitate to invest in research on the effects of their products on wellbeing, as negative findings could result in bad publicity as exemplified by the 2018 ‘Techlash’ at Facebook for instance \citep{hemphill_techlash_2019}. Further, the benefits of prioritizing user wellbeing over profits are currently ambiguous for technology companies. Specifically, companies grapple with uncertainty about whether prioritizing user wellbeing over profits, which often involves focusing more on ‘doing good’ rather than just ‘preventing harm,’ will yield tangible benefits \citep{morley_operationalising_2021}. This ambiguity is rooted in the difficulty of quantifying the return on investment for ethical AI practices that emphasize proactive welfare measures over mere harm avoidance. However, gaining clarity on this trade-off is impeded by corporate reluctance toward transparency and third-party algorithm audits, which are seen as jeopardizing competitive advantage \citep{stray_building_2023, stray_platforms_nodate}. This uncertainty, coupled with the perceived risks of optimizing for wellbeing, disincentivizes companies from allocating resources to human-centered design interventions. Overcoming these barriers will require establishing transparent accountability mechanisms, alternative business models not reliant on exploitation, and fostering an ethical culture recognizing that benefiting humanity and profits can be compatible \citep{di2020artificial}.

This reluctance highlights intricate tensions between public perception, economic incentives, and ethical duties facing corporate decision-makers. Such challenges of power, as also publicly displayed during the OpenAI-Altman debacle late 2023 \citep{ulanoff_openais_2023}, are highly relevant for Positive AI. These second-order challenges determine system-level goals, shaping whether wellbeing optimization occurs. If companies lack motivation to prioritize wellbeing, alignment is unlikely. Wellbeing as an overarching principle for AI alignment is promising but faces real-world obstacles regarding corporate priorities. While prominent voices endorse human flourishing as the goal, transparency and accountability mechanisms appear necessary to actualize this vision \citep{morley_ethics_2021}.

\subsection{Future opportunities}

AI benchmarking is a popular method for evaluating the capabilities of large language models (LLMs). As AI benchmarking matures and as AI permeates more aspects of life, more sophisticated will be benchmarks required \citep{burnell2023rethink}. For example, benchmarks for LLM qualities like toxicity are now widely used \citep{lynch2023ai}. Carefully crafted wellbeing metrics could serve as a mechanism for academics and others to indirectly optimize AI systems such as ChatGPT. This is because benchmarks are used to compare different models; as a result, low performance on a benchmark can motivate improvement. Human-centered designers may help attune evaluation methods such as benchmarking better to actual human experiences and ensure that optimization metrics align with these experiences.

\subsection{Limitations and final remarks}

While this paper aimed to provide an overview of key challenges in designing Positive AI, there are inherent limitations in addressing such a complex, transdisciplinary topic. For instance, environmental sustainability is not a core focus of the article, yet it warrants mention given the intricate relationship between sustainability and human wellbeing. As environmental crises increasingly threaten flourishing across communities, sustainability is being recognized as fundamental to comprehensive models of wellbeing \citep{kjell_sustainable_2011, omahony_toward_2022}. That is, wellbeing means more than human wellbeing. Meanwhile, the energy-intensive nature of AI systems presents sustainability challenges \citep{vinuesa_role_2020}. This surfaces an alignment tension between AI benefits and potential unintended harms. Additionally, many relevant issues, from philosophy of technology to data ethics, could only be briefly touched upon given the practical design focus of this paper, but present key areas for further investigation.

Nonetheless, this work aims to spark discussion and research at the intersection of AI, wellbeing, and design. Further interdisciplinary collaboration building on these ideas will develop more pluralistic perspectives on Positive AI. The goal of this work was to identify key challenges that must be addressed by designers in order to develop AI systems aligned with human wellbeing. As the fourth industrial revolution is well on its way, “the time of reckoning for artificial intelligence is now.” \citep{ozmen_garibay_six_2023}

\bibliographystyle{apacite}
\bibliography{references_diss, zotero}
\end{document}